# Impact of MvdW Equation of State and Neutrino Mass on r and s Process Heavy Element Nucleosynthesis in Spiral, Elliptical and Dwarf Galactic Environments and Kilonovae Events

Keith Andrew[1], Eric Steinfelds[1,2], Kristopher Andrew[3]

[1]Department of Physics and Astronomy, Western Kentucky University
Bowling Green, KY 42101 USA
[2]Department of Computational and Physical Sciences, Carroll University
Waukesha, WI 53186 USA
[3]Department of Science, Schlarman Academy,
Danville, IL 61832 USA

**Abstract:** We present an analysis of heavy element production with massive neutrinos in galaxies of varying types (spiral, elliptical, and dwarf) and kilonovae events by incorporating a Multicomponent Van der Waals (MvdW) equation of state (EoS) for the opacity functions. This EoS is applied to derive opacities and calculate the yields of isotopes formed in r-process and s-process nucleosynthesis, with and without the influence of neutrino masses or oscillations. We look at both the lanthanide and actinide sequences using the MvdW parameters that involve the interaction strength and excluded volume effects. Our results reflect the characteristic differences found in r and s processes in the synthesis and long-term evolution of isotopes from the U, Th, and Sr chain across galactic environments. The inclusion of neutrino masses enhances the neutron-to-proton ratio, favoring heavier r-process isotopes and altering the overall galactic yields by cross section suppression. These findings offer insights into the interplay of nuclear physics and astrophysical environments, highlighting the sensitivity of nucleosynthetic pathways to EoS modifications and neutrino physics. We compare these results to metallicity profiles of similar models: the Galactic Leaky Box, the Galactic Inflow, and the Galactic Closed Box models and to the kilonova event GW170781.

**I. Introduction** In this work we are exploring the use of an analytical phase changing equation of state (EoS), the multicomponent van der Waals (MvdW) EoS, to model the impact of neutrino masses or oscillations on heavy element production, mostly lanthanides and actinides, across galactic types: elliptical, spiral, and dwarf galaxies and a sample kilonova event similar to GW170817 as recorded by Swift, LIGO, and NuStar [1, 2, 3, 4]. While galactic and kilonova heavy element production methods differ in size and time scale, in this theoretical model they only differ by well-known mass consumption rate equations: the Galactic Star Formation Rate Equations [5, 6, 7] and the Kilonova Mass Ejection Rate Equation [8, 9, 10]. The production of heavy elements in the universe is governed by nucleosynthesis processes such as the rapid (r-) and slow (s-) neutron capture processes [11, 12], while there are other known mechanisms: such as the p-process or photodisintegration [13, 14], which can produce proton rich isotopes in Type-II core collapse supernovae and can produce rare elements, it is not a major contributor to galactic scale production, the i-process [15, 16] for the intermediate neutron flux case and occurs in He-shell flashes of asymptotic giant branch (AGB) stars, it is of limited significance on the galactic scale, the ν-process [17, 18, 19] or neutrino process is involved in neutrino induced nucleosynthesis via neutral and charged currents which usually produces lighter elements and requires a large neutrino flux not present here, the rp-process [20, 21] for rapid proton capture followed by beta decay which is most relevant in light element production and the f-process for extremely fast neutron capture, much faster than the r-process, which is considered very exotic and not impactful in our kilonova and galactic environments. These processes are influenced by astrophysical conditions and nuclear physics inputs, including the equation of state (EoS), opacity functions κ, cross sections σ, neutrino interactions, masses and/or oscillations, where we model using the neutrino upper bound value [22, 23], and galactic scale stellar environments. The early work by Bethe [24], followed by Alpher, Bethe, and Gamow [25], with the development of the r and s processes by Cameron [26], which were critical for the foundation of the work of Suess and Urey [27], and the detailed work by Margret Burbridge, Geoffrey Burbridge, William Fowler and Fred Hoyle [28], established the relationship between stellar nuclear reactions and the production of chemical elements after the Big Bang [29]. The path from light elements to isotopes of Fe is complex and involves an understanding of relativistic hydrodynamics, r process relativistic nuclear reaction networks [30], weak interactions and neutrino physics [31], cross sections [32] and high-density reaction rates [33] well beyond nuclear saturation number density near 0.16 fm$^{-3}$, energy transport with convection [34] and radiation, and numerous stable islands along with issues of stability [35] and energy release [$^{36}$]. Understanding these details often requires significant computational resources to focus on the best nuclear EoS. There are also several analytical models used to develop insightful and intuitive understanding of nuclear processes such as effective field theory [37], nuclear shell models [38], liquid drop models [39], lattice models [40], fission barrier and fission yield models [41], mean field theories [42], isospin theories [43], interacting plasma and fluid models [44, 45] and models based on phase changes similar to the modified and multicomponent van der Waals (MvdW) model [46, 47, 48, 49]. Once a star leaves the main sequence its future is intimately tied to its mass value. Many final states will involve mass transfer and mass ejections resulting in violent or unusual shock waves and perhaps free quark matter or a superconducting color flavor locked compact core [50, 51]. On the galactic scale the magnitude and distribution of the stellar birth rate and metallicity distribution and gradient varies for each galaxy type. The spirals exhibit relatively flat disk metallicity and constant stellar birth rates with higher core values indicating sustained r and s processes with compact binary mergers actively promoting events [52, 53], elliptical galaxies show enhanced core metallicity with early high r process yields with less metal-rich outskirts [54, 55], while dwarf galaxies show weak metallicity gradients and low metallicities with the potential for multiple starburst eras indicative of bursty and stochastic rates with long quiescent periods [56, 57]. These differences can be captured with the Galactic Star Formation Rate function dM/dt, for each galactic type.

Analytical modified equations of state can lead to an intuitive representation of dense astrophysical environments, such as neutron star mergers and core-collapse supernovae [58, 59]. Additionally, neutrino oscillations and masses affect the neutron-richness of these environments, altering the yields of r-process isotopes [60, 61, 62, 63, 64]. This study combines these factors to explore their effects on nucleosynthesis and isotopic abundances in different galaxy types and an example kilonova event using the MvdW EoS. The outline of the paper is as follows: in the next section we construct the theoretical model for elemental production, yields, opacities and EoS, then we look at the impact of varying the parameters of the EoS on element production, we then model galactic and kilonova production rates with and without massive neutrinos, we then compare these values to other models and observational data and finish with a conclusion.

## 2. Derivation of Opacities for MvdW EoS

The equation describing the balance between the production rate and loss mechanisms to determine the time rate of change of the abundance of a nuclear species X with number $N_X$, at any time is given as

$$\frac{dN_X}{dt} = \int \dot{M}_{star} \left[ Y_X(t) \Phi_{imf}(t) \right] dt - \left( N_X \kappa_{tot} \sigma T^4 + \lambda_X N_X \right)_{Loss}$$

$$\Phi_{imf} = \frac{1.35 M^{-2.35}}{M_{high}^{-1.35} - M_{low}^{-1.35}}$$

(1)

where we utilize the normalized Salpeter [65, 66, 67] initial mass function $\Phi_{imf}$ and the M overdot indicates the time derivative. The first term represents the production of the nuclear species X due to kilonovae stellar or galactic sources: for the kilonova example it is the mass rate from the ejecta, and for the galactic applications it is the stellar creation rate for each galaxy type, this is multiplied by the yield, Y, from fission decay or fusion capture processes for production, followed by the semiempirical Salpeter mass function, $\Phi_{imf}$, applied over the range of masses in the event for the highest to the lowest mass. From this we subtract the losses due to photodisintegration or loss due to neutron capture, which depends on the total opacity function $\kappa$, and the radioactive decay losses as the product of the decay rate $\lambda_x = \langle \sigma_x v \rangle$, with cross section $\sigma_x$, where we use the NIST thermal cross sections and rates [68] and BNL energy dependent cross sections with Legendre coefficients [69], and velocity v, and the number of nuclei $N_x$. The yield functions are given for the r process as

$$\left[ \frac{dY(Z,A)}{dt} \right]_r = \lambda_{n\gamma} Y(Z, A-1) - \lambda_{\gamma n} Y(Z, A) - \lambda_\beta Y(Z, A) + \lambda_{\beta^+} Y(Z-1, A)$$

$\lambda_{n\gamma}$ : Neutron-capture-Rate, $\quad \lambda_{\gamma n}$ : Photodisintegration-Rate

$\lambda_\beta$ : Beta-decay-Rate, $\quad\quad \lambda_{\beta^+}$ : Positron-Emission-Rate

(2)

and for the s process we have

$$\left[ \frac{dY(Z,A)}{dt} \right]_s = n_n \langle \sigma v \rangle (Z, A-1)_{cap} Y(Z, A-1) - n_n \langle \sigma v \rangle (Z, A) Y(Z, A)_{cap} + \lambda_{\beta^+}(Z-1, A) Y(Z-1, A) - \lambda_\beta(Z, A) Y(Z, A) \quad (3)$$

where $n_n$ is the neutron number density and $\sigma$ is the cross section, v is the velocity, A is the total mass number and Z is the proton atomic number. For the neutrino processes we have

$$\left[\frac{dY(Z,A)}{dt}\right]_\nu = n_\nu \left[\langle\sigma v\rangle_{\nu,abs}(Z-1,A)Y(Z-1,A) - \langle\sigma v\rangle_{\nu,abs}(Z,A)Y(Z,A) + \langle\sigma v\rangle_{\nu,cap}(Z,A)Y(Z,A)\right] \quad (4)$$

$$\nu_e + n \to p + e^- \qquad \bar{\nu}_e + p \to n + e^+ \quad .$$

The opacities can be represented as short wavelength UV or r process functions and long wavelength IR or s process functions where the wavelength dependence is critical for separating out the r and s processes for each term in the opacity expression: one for bound state to bound state electrons, bb, one for bound to free electron photoionization, bf, one for free-to-free electron Bremsstrahlung, ff, and the final term for electron scattering, es:

$$\kappa_\nu(a,b) = \left[\kappa_\nu^{bb}(a,b) + \kappa_\nu^{bf}(a,b)\right]_{UV,r} + \left[\kappa_\nu^{ff}(a,b) + \kappa_\nu^{es}(a,b)\right]_{IR,s} \quad (5)$$

where a and b are the MvdW EoS parameters. Each of these terms depends upon the EoS being used through the material density for each opacity term as

$$\kappa_\nu^{bb} = \sum_{i,j} \frac{\pi e^2}{m_e c} f_{ij} \frac{\rho}{m_u A} \phi_{ij}(\nu) \qquad \kappa_\nu^{bf} = \sum_i \frac{\rho}{m_u A} \sigma_i(\nu)$$

$$\kappa_\nu^{ff} = \frac{4e^2 Z^6}{3 m_e^2 c^3 h \nu^3} \left(\frac{2\pi}{3kT}\right)^{1/2} \left(\frac{\rho Y_e}{m_p}\right) \left(\frac{\rho}{m_u A}\right) g_{ff}(\nu) \qquad \kappa_\nu^{es} = \sigma_T \frac{\rho Y_e}{m_p}$$

$$\phi_{ij} = \phi(\nu - \nu_{ij}) = \frac{1}{\Delta\nu_D \sqrt{\pi}} \exp\left(-\frac{(\nu - \nu_{ij})^2}{\Delta\nu_D^2}\right), \qquad \Delta\nu_D^2 = \left(\frac{2kT}{m_e c^2}\right) \nu_{ij}^2$$

$$g_{ff} = 1.1 \qquad 0.04 < f_{ij} < 0.93$$

(6)

where we use the photoionization free-free Gaunt-Kramers [70] astrophysical factor of $g_{ff} = 1.1$, and the oscillator strengths, $f_{ij}$, are from the NIST database [71, 72], and $\varphi_{ij}$ is the transition line shape for frequency $\nu_{ij}$. For the MvdW EoS the density function can be solved directly in terms of temperature and pressure. The MvdW EoS, has a partition function expressed as:

$$Z_{vdw}(N_i, V, T) = \prod_{i=1}^{N_C} \frac{1}{N_i!} \left(\frac{V - \sum_{j=1}^{N_C} N_j b_j}{\Lambda_i^3}\right)^{N_i} \exp\left(\frac{N_i}{VkT} \sum_{j=1}^{N_C} a_{ij} N_j\right)$$

(7)

$$\Lambda_i = \sqrt{\frac{1}{2\pi m_i T}} \qquad a_{ij} = \sqrt{a_i a_j}\,(1 - k_{ij})$$

where $\Lambda_i$ is the thermal deBroglie wavelength and the $a_{ij}$ represent the interparticle interaction strength, which is symmetric, $a_{ij}=a_{ji}$, and $a_{ii}=a_i$ and $k_{ij}$ is the standard mixing parameter between $a_i$ and $a_j$ and the $b_j$ represent the excluded volume factors and the $a_{ij}$ strength is determined by the QCD color factor for the representation of $SU(3)_c$ being considered, where the effective $q_i q_j \delta_{ij}$ quark interaction is attractive and repulsive [73]. This gives the MvdW EoS as

$$p = T\left(\frac{\partial \ln Z}{\partial V}\right)_{N,T} = T\sum_{i=1}^{N_c}\left[\frac{N_i}{V - \sum_{j=1}^{N_c} N_j b_j} - \frac{N_i}{V^2 T}\sum_{j=1}^{N_c} N_j a_{ij}\right] = \sum_{i=1}^{N_c}\left[\frac{\rho_i T}{1 - \sum_{j=1}^{N_c} \rho_j b_j} - \sum_{j=1}^{N_c} \rho_i \rho_j a_{ij}\right] \quad (8)$$

which may be expressed in terms of particle number or density factors. For the kilonova and galactic mass rates we use the semiempirical models using a range of mass values to match observations and the three standard galactic functions to match spiral, elliptical, and dwarf galaxies [74,75].

Galactic Star Formation Rates

$Sp: \dot{M}_{star}(t) = \dot{M}_0 \quad 1 < \dot{M}_0 < 10 \, M_{solar}/yr$

$El: \dot{M}_{star}(t) = \dot{M}_0 e^{-t/\tau} \quad 0.5 < \tau < 1 \, Gyr$

$Dw: \dot{M}_{star}(t) = \sum_i \dot{M}_i e^{-(t-t_i)^2/2\sigma^2} \quad 0.01 < \sigma < 0.1 \, Gyr$

$KN - Mass - Ejecta - Rate$

$\dot{M}_{kn} = \frac{M_{ejecta}}{\tau_{ejecta}\gamma_v} \quad 10^{-4} < M < 0.1 \, M_{solar}$

$\tau_{ejecta} \sim \frac{R_{ejecta}}{v_{ejecta}} \quad \gamma_v = (1-v^2)^{-1/2}, R_{ejecta} \sim 10^{10} \, cm$

$v_{ejecta} \sim 0.2, \quad \gamma_v = 1.04$

(9)

where $\gamma_v$ is the relativistic Lorentz factor. Here we have established a direct link between the MvdW EoS parameters and nucleosynthetic yields by modifying opacity equations and establishing a model that can be used to explore the element production rate differences with and without massive neutrinos. The neutrino energies, and mass values, are in the cross sections needed for each rate value $\lambda_i$.

## 3. KN and Galactic Results

The r process is a rapid neutron capture that occurs when the neutron flux is high and the capture is faster than the decay rate thereby producing a neutron rich environment that can be far from the valley of stability. In the actinides (89 < Z < 103) production of $^{232}$Th, and $^{238}$U occurs, with a peak near A~130, in the lanthanides (57 < Z < 71) $^{139}$La, $^{150}$Sm, and $^{152}$Eu are produced with a peak near A~87. The s process dominates in low to moderate neutron flux regions capturing neutrons slowly so that beta decay may occur between captures. The details of the reaction chain used here are given in Appendix A. In Fig. 1 we show the production rates for light and heavy elements as the MvdW EoS parameters are varied for a kilonova event.

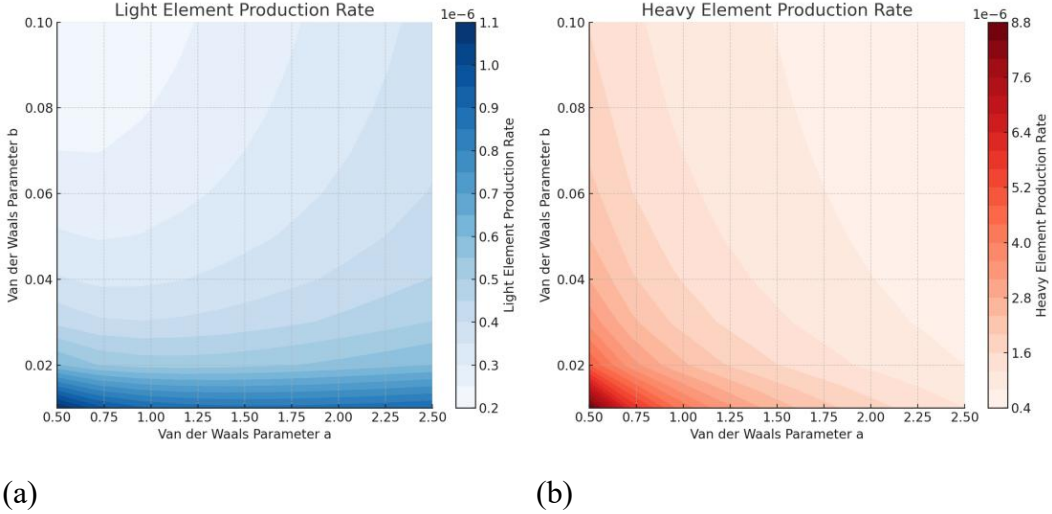

(a)                                          (b)

Fig. 1 In (a) the light element production rate is shown as a function of the two MvdW EoS parameters a and b are varied and in (b) we show the heavy element production rate as the MvdW parameters are again varied: 0.5 < a < 2.5, and 0.02 < b < 0.1 . The heavy element production rate is more sensitive to both a and b favoring the smaller values for higher production.

We next consider an r process and s process production sequence shown in Fig 2. Here we have the r process rates for: $^{56}$Fe, $^{56}$Co, $^{56}$Ni, $^{56}$Zn and s process rates for: $^{56}$Fe, $^{57}$Fe, $^{58}$Fe, $^{88}$Sr, $^{89}$Sr, $^{90}$Sr, $^{138}$Ba, $^{139}$Ba for a kilonova event as a function of time, with and without massive neutrinos. The massive neutrino effects can be seen from the massive neutrino cross section suppression effect. Essentially adding the mass term to the full neutrino 4-momentum reduces the available reaction energy and changes the cross section as can be seen from the ratio of cross sections in Eq.(10). These cross sections are used to determine the rate factors λ causing a change in the overall production rate.

$$\frac{\sigma_{m_\nu}}{\sigma_{m_\nu=0}} = \left(\frac{\sqrt{E_\nu^2 - m_\nu^2}}{E_\nu}\right)^2 \tag{10}$$

In our model the ejecta mass is set to 0.1 M$_\odot$ and both neutrino cases are compared where we have used solid curves for the massless case and dashed curves for the massive case and a value of 0.8 eV for the electron neutrino mass.

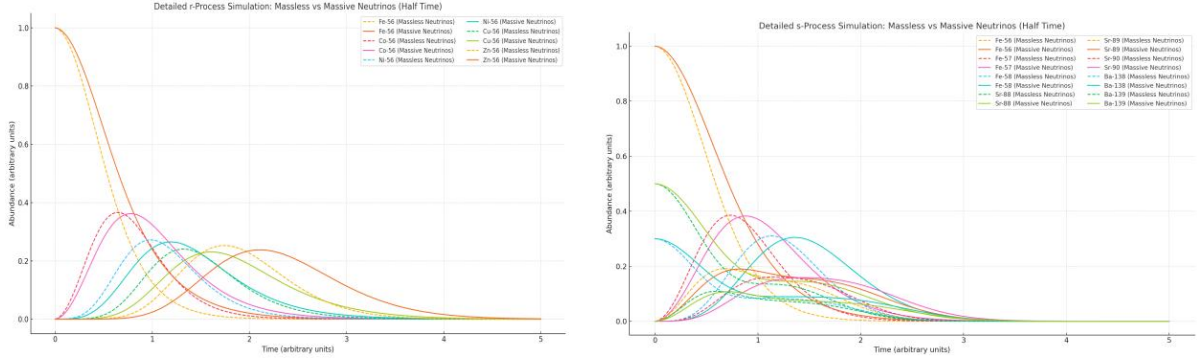

Fig. 2 The r and s process production for a kilonova event, in (a) Fe, Co, Ni and Zn isotopes are shown for the r-process and in (b) Fe, Co, Sr, and Ni isotopes are shown for the s-process.

For the galactic case we use the galactic star formation rates given in Eq. (9) which are designed to reflect the physical and observational properties of each galaxy type. For the spiral galaxies we have a steady inflow of gas from the intergalactic medium or recycling from stellar feedback. The constant stellar formation rate represents a reasonable and sustainable mass supply over billions of years. For elliptical galaxies there is an early intense starburst period consuming and then expelling gas followed by a period of quiescence leading to a consistent fit with exponential decay. For dwarf galaxies there is a more stochastic star formation rate that can occur on different time scales, so each term in the series represents a different active era as seen in such low mass galaxies. In Fig. (3) the heavy element abundances are shown as a function of galactic type and with and without massive neutrinos.

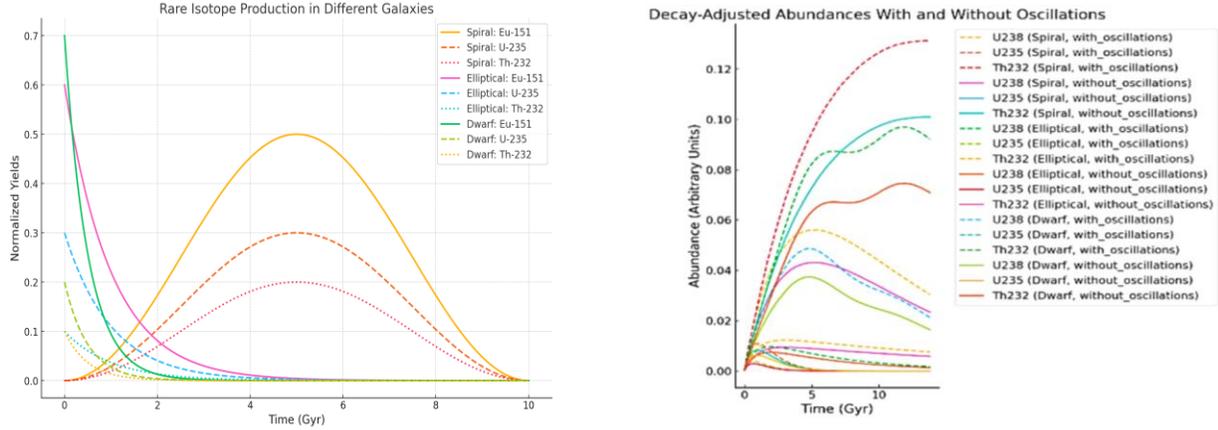

Fig. 3. In (a) we show the galactic yields for isotopes of U, Eu and Th for each galactic type, elliptical, spiral and dwarf and in (b) we show the galactic normalized abundances with and without massive neutrinos which induce oscillations.

In Fig. 4 we first show the abundances in two kilonovae, one with a 1.2 $M_\odot$ ejecta and one with a 1.5 $M_\odot$ ejecta with and without neutrinos. In both cases the massive neutrinos reduce the r-process yields thereby reducing the heavy element abundance while the higher ejecta mass value contributes more to the enrichment process.

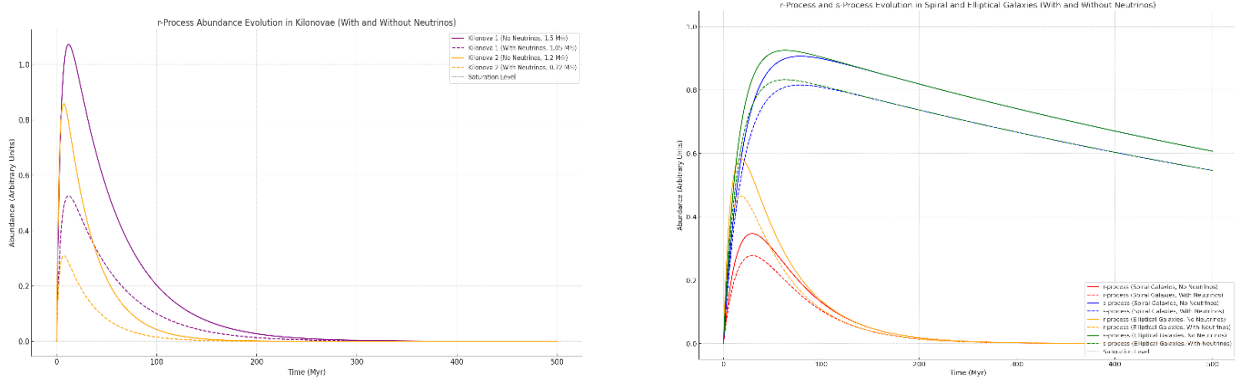

Fig. 4 (a) Abundances for two kilonovae, one with 1.2 M$_\odot$ ejecta and one for 1.5 M$_\odot$ ejecta, (b) r-process and s-process abundances for elliptical and spiral galaxies with and without neutrino masses.

In Fig. 4 (b) we show the galactic abundances for the r-process and s-process yields for elliptical and spiral galaxies with and without neutrinos. The elliptical galaxies show a rapid early enrichment phase, especially for the r-process in the first Gyr, followed by a steady decline. The neutrino masses significantly reduce the peak values and early production which is followed by a rapid decay. The s-process is similar and slightly slower with a smaller reduction due to the neutrino masses. In the spiral galaxies the enrichment is slower but richer with a higher peak followed by a much slower decline, where again the massive neutrinos reduce the overall production at peak values, and significantly at later times.

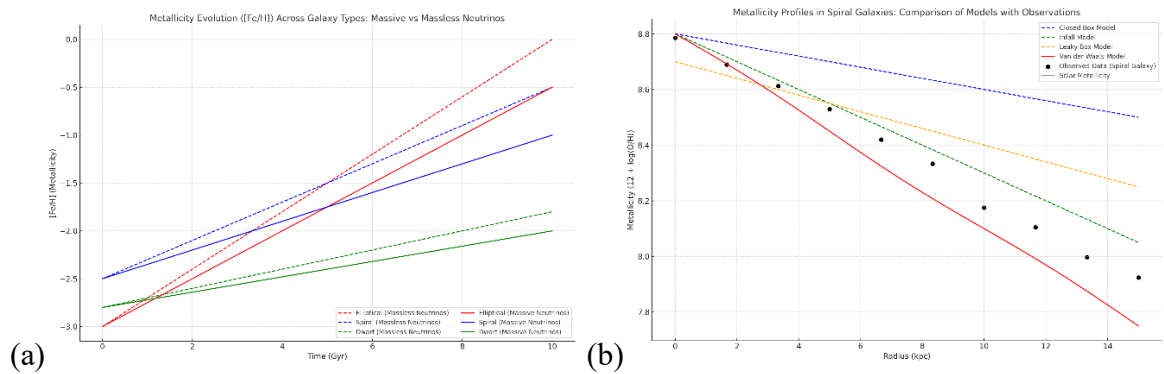

(a) (b)

Fig. 5 In (a) we show the metallicity as a function of time for each galactic type with and without massive neutrinos and in (b) we compare the MvdW massive neutrino model to an observational data set and similar metallicity calculations.

In Fig (5) we analyze metallicity [Fe/H] evolution across galaxy types with and without massive neutrinos. This comparison demonstrates the impact of neutrino masses causing slower metal production and delaying enrichment for each galaxy type where the most pronounced effect is for spiral galaxies. In all cases, when compared to observational data, the massive neutrino model shifts closer to the observational data. The MvdW EoS model is also compared to three other similar models and to an observational data set. The Leaky Box Model [76, 77], which accounts for gas outflows, where metals are ejected from the galaxy due to supernovae driven winds. The Infall Model [78, 79, 80] incorporates gas inflow, where pristine or metal poor gas dilutes the galaxy's metallicity and sustains star formation. The Closed Box Model [81, 82] assumes negligible gas inflow or outflow, metals accumulate over time from stellar evolution inside the galaxy. This is compared to the averaged observational data for metallicity curves of spiral galaxies [83, 84], principally from the Milky Way [85, 86] and M81 (NGC 3031) [87] which is a nearby (11.8 Mly) grand design spiral.

## 4. Conclusion

The use of the MvdW EoS provides a framework for examining the production of heavy elements in diverse galactic environments. This EoS captures critical physical effects in high-density, high-temperature conditions, such as those found in neutron star mergers, kilonovae events, and supernovae ejecta. By integrating this EoS into opacity expressions, we achieve an intuitive representation of photon and neutrino interactions with matter, which directly influence nucleosynthesis pathways in the r and s processes.

Our analysis demonstrates that galaxy types exhibit distinct heavy element production profiles that are impacted by neutrino masses: spiral galaxies sustain steady heavy element production due to their continuous star formation histories, leading to cumulative enrichment of both r- and s-process isotopes, elliptical galaxies show a rapid early peak in heavy element yields, dominated by their initial starburst, followed by a plateau as star formation ceases, dwarf galaxies exhibit episodic production, reflecting their bursty star formation and feedback cycles, with pronounced variability in r-process yields.

Neutrino masses or oscillations play a pivotal role in shaping r-process yields by enhancing neutron-rich conditions, particularly in high-density environments. This effect is most significant in spiral galaxies where the new star production rate is active over a long period enhancing r-process production at the end of the stellar life cycle. Neutrino oscillations play a crucial role in determining the yields of heavy elements by altering the neutron-richness of astrophysical environments, shifting the balance between light and heavy isotopes, profoundly impacting the chemical evolution of galaxies and the universe. As such massive neutrino physics is important for accurately modeling the synthesis and evolution of heavy elements. These results have significant implications for understanding the chemical evolution of galaxies and the cosmic distribution of elements, highlighting the importance of combining detailed nuclear physics with astrophysical simulations.

This study demonstrates the critical role of the MvdW EoS parameters (a, b) and neutrino mass values in shaping r- and s-process nucleosynthesis across galaxy types and in kilonovae. High $b_i$ values more closely describe the r-process dominance in the galaxy like Reticulum II [88, 89], while moderate $a_{ij}$ balances spiral galaxy heavy element abundances. Neutrino masses suppress neutron cross sections, reducing elemental abundances in all galactic types and delay heavy element nucleosynthesis. These findings align theoretical predictions based on using the MvdW

EoS used in a microscopic setting to derive r and s process opacities, including terms for massive neutrinos, with observations across galactic types and kilonovae events, providing insights into the origins of heavy element production and an understanding of the physics of extreme matter.

---

33. Klähn, T., D. Blaschke, S. Typel, E. N. E. Van Dalen, A. Faessler, C. Fuchs, T. Gaitanos et al. "Constraints on the high-density nuclear equation of state from the phenomenology of compact stars and heavy-ion collisions." *Physical Review C—Nuclear Physics* 74, no. 3 (2006): 035802.
34. Mathis, Stéphane. "Variation of tidal dissipation in the convective envelope of low-mass stars along their evolution." *Astronomy & Astrophysics* 580 (2015): L3.
35. Murphy, George L. "Do superheavies come from neutron stars?." *Nature* 263, no. 5573 (1976): 114-115.
36. Clayton, Donald D. *Principles of stellar evolution and nucleosynthesis*. University of Chicago press, 1983.
37. Rauscher, Thomas, A. Heger, R. D. Hoffman, and S. E. Woosley. "Nucleosynthesis in massive stars with improved nuclear and stellar physics." *The Astrophysical Journal* 576, no. 1 (2002): 323.
38. Martínez-Pinedo, Gabriel, and Karlheinz Langanke. "Shell model applications in nuclear astrophysics." *Physics* 4, no. 2 (2022): 677-689.
39. Furusawa, Shun, Kohsuke Sumiyoshi, Shoichi Yamada, and Hideyuki Suzuki. "New equations of state based on the liquid drop model of heavy nuclei and quantum approach to light nuclei for core-collapse supernova simulations." *The Astrophysical Journal* 772, no. 2 (2013): 95.
40. Lee, Dean, Buḡra Borasoy, and Thomas Schaefer. "Nuclear lattice simulations with chiral effective field theory." *Physical Review C—Nuclear Physics* 70, no. 1 (2004): 014007.
41. Goriely, Stéphane. "The fundamental role of fission during r-process nucleosynthesis in neutron star mergers." *The European Physical Journal A* 51 (2015): 1-21.
42. Li, A., Z-Y. Zhu, E-P. Zhou, J-M. Dong, J-N. Hu, and C-J. Xia. "Neutron star equation of state: Quark mean-field (QMF) modeling and applications." *Journal of High Energy Astrophysics* 28 (2020): 19-46.
43. Sharma, Bharat K., and Subrata Pal. "Role of isospin physics in supernova matter and neutron stars." *Physical Review C—Nuclear Physics* 82, no. 5 (2010): 055802.
44. Andrew, Keith, Eric V. Steinfelds, and Kristopher A. Andrew. "Cold Quark–Gluon Plasma EOS Applied to a Magnetically Deformed Quark Star with an Anomalous Magnetic Moment." *Universe* 8, no. 7 (2022): 353.
45. Spartà, Roberta, M. La Cognata, G. L. Guardo, S. Palmerini, M. L. Sergi, G. D'Agata, L. Lamia et al. "Neutron-Driven Nucleosynthesis in Stellar Plasma." *Frontiers in Physics* 10 (2022): 896011.
46. Malaver, Manuel, and Hamed Daei Kasmaei. "Analytical models for quark stars with van der Waals modified equation of state." *International Journal of Astrophysics and Space Science* 7, no. 5 (2019): 58.
47. Andrew, Keith, Eric V. Steinfelds, and Kristopher A. Andrew. "The van der Waals Hexaquark Chemical Potential in Dense Stellar Matter." *Particles* 6, no. 2 (2023): 556-567.
48. Malaver, Manuel. "Analytical model for charged polytropic stars with Van der Waals Modified Equation of State." *American Journal of Astronomy and Astrophysics* 1, no. 4 (2013): 41-46.
49. Andrew, Keith, Eric Steinfelds, and Kristopher Andrew. "Signature of the Gravity Wave Phase Shift in a Cold Quark Star with a Nonconvex Multicomponent van der Waals Equation of State." *arXiv preprint arXiv:2412.07601* (2024).
12

Appendix A
Reaction Sub-chain Equations

To calculate the yields for the entire lanthanide or actinide r-process or s-process would require a coupled system of some 765 differential equations. In order to work with a more tractable reaction chain that still captures the overall pattern we select sub-chains that focus on a representative sample of each process of interest. The resulting systems used in each figure for each process are given below along with the requisite seed nuclei to start the chain form the stellar core.

The notation for the rate values is given as:

$$\text{Neutron number density: } n_n$$
$$\text{Reaction rate for neutron capture: } \langle \sigma v \rangle_{Z,A}$$
$$\text{Photodisintigration rate: } \lambda_{\gamma-n}(Z, A)$$
$$\text{Beta decay rate: } \lambda_{\beta}(Z, A)$$
$$\text{Spontaneous fission rate: } \lambda_{sf}(Z, A)$$

The lanthanide seeds and sub-chain for Fig.(2)(a) r-process is:

$^{56}$Fe seed:
$$\frac{dY(26,56)}{dt} = -n_n \langle \sigma v \rangle_{26,56} Y(26,56) + \lambda_{\gamma-n}(26,57) Y(26,57)$$

$^{59}$Co seed:
$$\frac{dY(27,59)}{dt} = -n_n \langle \sigma v \rangle_{27,59} Y(27,59) + \lambda_{\gamma-n}(27,60) Y(27,60)$$

$^{64}$Ni seed:
$$\frac{dY(28,64)}{dt} = -n_n \langle \sigma v \rangle_{28,64} Y(26,56) + \lambda_{\gamma-n}(28,65) Y(28,65)$$

$^{68}$Zn seed:
$$\frac{dY(30,68)}{dt} = -n_n \langle \sigma v \rangle_{30,68} Y(30,68) + \lambda_{\gamma-n}(30,69) Y(30,69)$$

$^{56}Fe : {}^{56}Co : {}^{56}Cu : {}^{56}Zn$

$$\frac{dY(26,56)}{dt} = -n_n \langle\sigma v\rangle_{26,56} Y(26,56) + \lambda_{\gamma-n}(26,57) Y(26,57)$$

$$\frac{dY(27,56)}{dt} = -n_n \langle\sigma v\rangle_{27,56} Y(27,56) + \lambda_\beta(26,56) Y(26,56) + \lambda_{\gamma-n}(27,57) Y(27,57)$$

$$\frac{dY(29,56)}{dt} = -n_n \langle\sigma v\rangle_{29,56} Y(29,56) + \lambda_\beta(28,56) Y(28,56) + \lambda_{\gamma-n}(29,57) Y(29,57)$$

$$\frac{dY(30,56)}{dt} = -n_n \langle\sigma v\rangle_{30,56} Y(30,56) + \lambda_\beta(29,56) Y(29,56) + \lambda_{\gamma-n}(30,57) Y(30,57)$$
.

The sub-chain s-process for Fig.(2)(b) is:

$^{56}Fe : {}^{57}Fe : {}^{58}Fe : {}^{88}Sr : {}^{89}Sr : {}^{138}Ba : {}^{139}Ba$

$$\frac{dY(26,56)}{dt} = -n_n \langle\sigma v\rangle_{26,56} Y(26,56) + \lambda_\beta(25,56) Y(25,56)$$

$$\frac{dY(26,57)}{dt} = n_n \langle\sigma v\rangle_{26,56} Y(26,56) - n\langle\sigma v\rangle_{26,57}(26,57) Y(26,57)$$

$$\frac{dY(26,58)}{dt} = n_n \langle\sigma v\rangle_{26,57} Y(26,57) - n\langle\sigma v\rangle_{26,58}(26,58) Y(26,58)$$

$$\frac{dY(38,88)}{dt} = -n_n \langle\sigma v\rangle_{38,88} Y(38,88) + \lambda_\beta(37,88) Y(37,88)$$

$$\frac{dY(38,89)}{dt} = n_n \langle\sigma v\rangle_{38,88} Y(38,88) - n\langle\sigma v\rangle_{38,89}(38,89) Y(38,89)$$

$$\frac{dY(38,90)}{dt} = n_n \langle\sigma v\rangle_{38,89} Y(38,89) - n\langle\sigma v\rangle_{38,90}(38,90) Y(38,90)$$

$$\frac{dY(56,138)}{dt} = -n_n \langle\sigma v\rangle_{56,138} Y(56,138) + \lambda_\beta(55,138) Y(55,138)$$

$$\frac{dY(56,139)}{dt} = n_n \langle\sigma v\rangle_{56,138} Y(56,138) - n_n \langle\sigma v\rangle_{56,139}(56,139) Y(56,139)$$
.

The sub-chain for Fig.(3)(a) r-process is:

$^{151}Eu : {}^{232}Th : {}^{235}U : {}^{238}U$

$$\frac{dY(63,151)}{dt} = n_n \langle \sigma v \rangle_{63,150} Y(63,150) - n_n \langle \sigma v \rangle_{63,151} Y(63,151) + \lambda_\beta (62,151) Y(62,151) - \lambda_\beta (63,151) Y(63,151)$$

$$\frac{dY(90,232)}{dt} = n_n \langle \sigma v \rangle_{90,231} Y(90,231) - n_n \langle \sigma v \rangle_{90,232} Y(90,232) + \lambda_\beta (89,232) Y(89,232) - \lambda_{sf} (90,232) Y(90,232)$$

$$\frac{dY(92,235)}{dt} = n_n \langle \sigma v \rangle_{92,234} Y(92,234) - n_n \langle \sigma v \rangle_{92,235} Y(92,235) + \lambda_\beta (91,235) Y(91,235) - \lambda_{sf} (92,235) Y(92,235)$$

$$\frac{dY(92,238)}{dt} = n_n \langle \sigma v \rangle_{92,237} Y(92,237) - n_n \langle \sigma v \rangle_{92,238} Y(92,238) + \lambda_\beta (91,238) Y(91,238) - \lambda_{sf} (92,238) Y(92,238).$$

The sub-chain for Fig(3)(b) r-process is given as:

$^{232}Th : {}^{235}U : {}^{238}U : {}^{244}Pu$

$$\frac{dY(90,232)}{dt} = n_n \langle \sigma v \rangle_{90,231} Y(90,231) - n_n \langle \sigma v \rangle_{90,232} Y(90,232) + \lambda_\beta (89,232) Y(89,232)$$

$$\frac{dY(92,235)}{dt} = n_n \langle \sigma v \rangle_{92,234} Y(92,234) - n_n \langle \sigma v \rangle_{92,235} Y(92,235) + \lambda_\beta (91,235) Y(91,235) - \lambda_{sf} (92,235) Y(92,235)$$

$$\frac{dY(92,238)}{dt} = n_n \langle \sigma v \rangle_{92,237} Y(92,237) - n_n \langle \sigma v \rangle_{92,238} Y(92,238) + \lambda_\beta (91,238) Y(91,238) - \lambda_{sf} (92,238) Y(92,238)$$

$$\frac{dY(94,244)}{dt} = n_n \langle \sigma v \rangle_{94,243} Y(94,243) - n_n \langle \sigma v \rangle_{94,244} Y(94,244) + \lambda_\beta (93,244) Y(93,244) - \lambda_{sf} (94,244) Y(94,244).$$